\documentclass[aps,prl,showpacs,twocolumn]{revtex4}
\usepackage{amsmath,graphicx,epsfig,amssymb,subfigure,times,dsfont}
\usepackage[usenames]{color}
\usepackage{bbold}

\newcommand {\Ref}[1] {Ref.~\cite{#1}}
\newcommand {\Fig}[1] {Fig.~\ref{#1}}
\newcommand {\Eqn}[1] {Eq.~(\ref{#1})}

\newcommand {\SectionCommand}[1] {\paragraph{#1:}}

\begin{document}

\title{Fourier transform of fermionic systems and the spectral tensor network}

\author{Andrew~J.~Ferris}
\affiliation{D\'epartement de Physique, Universit\'e de Sherbrooke, Qu\'ebec, J1K 2R1, Canada}
\affiliation{ICFO---Institut de Ciencies Fotoniques, Parc Mediterrani de la Tecnologia, 08860 Barcelona, Spain}
\affiliation{Max-Planck-Institut f\"ur Quantenoptik, Hans-Kopfermann-Str. 1, 85748 Garching, Germany}

\date{\today}

\begin{abstract}
Leveraging the decomposability of the fast Fourier transform, I propose a new class of tensor network that is efficiently contractible and able to represent many-body systems with local entanglement that is greater than the area law. Translationally invariant systems of free fermions in arbitrary dimensions as well as 1D systems solved by the Jordan-Wigner transformation are shown to be exactly represented in this class. Further, it is proposed that these tensor networks be used as generic structures to variationally describe more complicated systems, such as interacting fermions. This class shares some similarities with Evenbly \& Vidal's branching MERA, but with some important differences and greatly reduced computational demands.
\end{abstract}


\maketitle


Solving strongly-correlated many-body systems is one of the primary challenges of modern quantum mechanics that arises in a wide variety of fields and contexts. Because direct approaches are usually intractable, an array of analytical and numerical techniques and approximations have been developed over the last century, each associated with its own region of validity, accuracy and difficulty.

One relatively modern approach is based on tensor network decompositions of many-body wavefunctions~\cite{Verstraete2008,Cirac2009,Orus2013,Eisert2013}, motivated by our understanding of quantum information in this context. By decomposing the wavefunction into a particular product of tensors, one defines the structure of entanglement that is permitted (for example, the low-energy states of many local Hamiltonians obey an \emph{area law} scaling for entanglement entropy~\cite{Hastings2007b,Wolf2008,Masanes2009,Eisert2010}). Typically, one applies the variational principle within this class of wavefunctions to determine, for example, the tensor network state closest to the ground state. Increasing the rank of the decomposition increases the amount of allowable entanglement, and potentially leads to a more accurate description of the many-body state.

A wide variety of tensor network geometries have been proposed, beginning with the matrix-product state (MPS)~\cite{Fannes1992,Oestlund1995a,Vidal2003} that is the foundation of the famous density matrix renormalization group~\cite{White1992} and Wilson's numerical renormalisation group~\cite{Wilson1975}. This one-dimensional structure connects nearest neighbors in a chain, and is extremely accurate for 1D systems that have spatially-local correlations (proven to be exponentially close to the low-energy states of gapped, local 1D Hamiltonians~\cite{Hastings2007b}). Real-space renormalization group ideas lead to the development of the tree-tensor network (TTN)~\cite{Shi2006,Tagliacozzo2009} and the multi-scale entanglement renormalization ansatz (MERA)~\cite{Vidal2007b,Vidal2008}. The MERA in particular is able to reproduce larger amounts of entanglement such as that necessary to describe critical, 1D quantum systems --- and has been found to correctly reproduce the features of the corresponding universality class at the renormalization fixed-point, as described by conformal field theory~\cite{Pfiefer2009}.

Extensions to higher dimensional systems are also possible. Although the MPS and TTN do not have the correct geometry to represent the area law for large 2D systems~\cite{Stoudenmire2012,Tagliacozzo2009,Ferris2013}, the MPS can be generalized to to a two-dimensional tensor network state commonly referred to as a projected entangled-pair state (PEPS)~\cite{Verstraete2004,Jordan2008,Gu2008}. The MERA is also generalized to higher dimensions in a straightforward fashion~\cite{Evenbly2009}. All the above techniques have shown success in solving 2D systems (e.g. \cite{Murg2007, Corboz2011,Yan2011,Harada2012}), but are somewhat handicapped by relatively large computational demands for even slightly-entangled systems.

Furthermore, both PEPS and the 2D MERA are limited to obey the area-law for entanglement entropy, where the entanglement entropy of a region is proportional to its boundary area (or perimeter, in 2D). This excludes important classes of physical systems which violate the area law --- notably any 2D system exhibiting a 1D Fermi surface (resulting in a logarithmic violation~\cite{Gioev2006}), and exotic phases such as Bose metals~\cite{Sheng2009,Zhang2011,Lai2013}.

This situation left the field of tensor networks in an awkward position, unable to accurately deal with what many outside the field consider to be the simplest models of non-interacting free-fermions, and their perturbations (i.e. Fermi liquids). More recently, a new class of tensor network that exhibits very large amounts of entanglement (up to volumetric) was proposed by Evenbly \& Vidal, which they have dubbed the `branching MERA'~\cite{Evenbly2012}. However, its properties and relationship to Fermi liquids has not been determined, as far as the author is aware.

In this work, I propose a new tensor-network geometry that mimics the structure of a spectral transformation and includes the quantum Fourier transform (for fermions) as a trivial example. Up to a reordering of the lattice, this `spectral tensor network' can be seen to be a generalization of TTN, or equally a simplification of branching MERA, and local expectation values can be extracted from a wavefunction of $n$ lattice sites in time $n \log n$. Similarly, it is related to previously suggested circuits for use in linear-optics quantum computing~\cite{Barak2007} and its fermionic analogue~\cite{Verstraete2009}. Here, I will give exact wavefunctions for translationally-invariant free fermion systems in arbitrary dimensions as well as examples of 1D systems that are exactly solvable via Jordan-Wigner transformations (Heisenberg XX chains, transverse-field Ising model, etc). I will then describe a procedure for finding a variational approximation of ground state of arbitrary local Hamiltonians in an efficient manner, which may be particularly relevant to simulations of 2D systems in the Fermi liquid phase, before concluding with remarks on future directions.

\SectionCommand{Quantum fast Fourier transforms}

We now discuss the unitary circuit required to transform a quantum lattice system between real- and momentum-space, and write it as a log-depth circuit using at most two-body gates. The discrete Fourier transform has long been known to  decompose into a series of sparse operations~\cite{Danielson1942, Cooley1965}, in a technique generally referred to as the fast Fourier transform (FFT). Here we will leverage this decomposition to describe the unitary circuit for the quantum Fourier transform for fermionic lattice models (similar to \cite{Verstraete2009}), and in the next section show that local expectation values of wavefunctions based on this circuit can be efficiently determined. 

Following the spirit of Danielson and Lanczos~\cite{Danielson1942}, we first demonstrate that the fermionic Fourier transform over $n$ (even) sites can be decomposed into two parallel Fourier transforms over $n/2$ sites followed by simple 2-body linear gates (matchgates):
\begin{equation}
  \sum_{x=0}^{n-1} e^{\frac{2 \pi i k x}{n}} \hat{c}^{\dag}_x = \!\sum_{x^{\prime} = 0}^{n/2-1} \!e^{\frac{2 \pi i k x^{\prime}}{n/2}} \hat{c}^{\dag}_{2x^{\prime}} + e^{\frac{2 \pi i k}{n}} \!\sum_{x^{\prime} = 0}^{n/2-1} \!e^{\frac{2 \pi i k x^{\prime}}{n/2}} \hat{c}^{\dag}_{2x^{\prime}+1} \label{decompose}
\end{equation}
The fermionic creation operator for site $x$ is denoted by $\hat{c}^{\dag}_x$, and the LHS of the above is (proportional to) the creation operator for the momentum mode $k$ (with wave-number $2\pi k / n)$. The RHS is the sum of two terms, each corresponding to a Fourier-transformed mode on the subset of even (or odd) sites. This sum is a linear transformation between two modes that can be implemented with a `beam-splitter' and `phase-delay' (graphically depicted in \Fig{decompose}). The beam-splitter is nothing other than the two-site Fourier transform
\begin{equation}
    \hat{F}_2 = \left[ \begin{array}{cccc} 1 & 0 & 0 & 0 \\ 0 & 2^{-1/2} & 2^{-1/2} & 0 \\ 0 & 2^{-1/2} & -2^{-1/2} & 0 \\ 0 & 0 & 0 & -1 \end{array} \right],
\end{equation}
written above in the number basis $\{|00\rangle,|01\rangle,|10\rangle,|11\rangle\}$. Notice that $\hat{F}_2 |1 1\rangle = -|1 1\rangle$ because of the anti-commutation relations of the creation/annihilation operators. The phase-delay takes care of the so-called twiddle-factor $e^{2\pi i k/n}$, and is written $\hat{\omega}^k_n = \hat{Z}^{2k/n}$ ($\hat{Z} = \hat{\mathbb{1}} - 2 \hat{n}$ is the Pauli matrix, $\hat{n}$ being the occupation operator).

\begin{figure}[t!]
\begin{centering}
  \includegraphics[width=1\columnwidth]{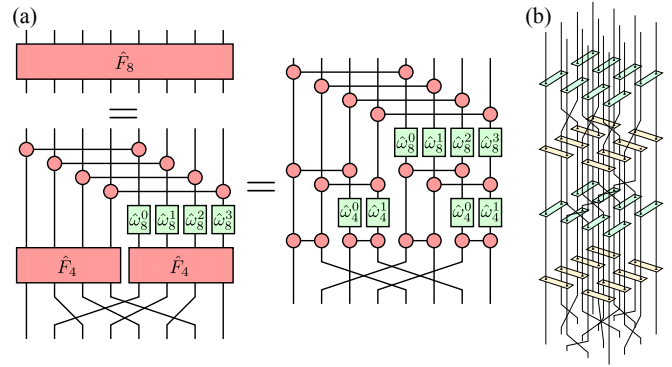}
  \caption{(Color online) (a) The unitary circuit for performing the quantum Fourier transform on 8 fermionic sites, $\hat{F}_8$, can be decomposed into 1- and 2-body gates using \Eqn{decompose}. The connected circular red gates are the two-site Fourier transform, $\hat{F}_2$. Real space is at the bottom of the diagram and momentum space at the top. (b) The decomposition also works in 2D, where here the 1-body gates are absorbed into the 2-body gates (which are therefore distinct). \label{fig:QFT}}
\end{centering}
\end{figure}

The decomposition in \Eqn{decompose} can be iterated $k$ times to fully determine the Fourier transform over $2^k$ sites --- however one must be careful to keep track of the ordering of modes, given by a bit-reversal operation. The resulting circuit for 8 sites is given in \Fig{fig:QFT}~(a). The structure of the circuit is identical to the Cooley-Tukey FFT~\cite{Cooley1965}. One can independently verify that the circuit produces the correct transformation because it is (a) linear in the field operators, and (b) performs the correct transformation on the single-particle subspace, in analogy with the classical FFT.

For systems of higher dimensions, the Fourier transform along each dimension can be applied sequentially along parallel wires and then decomposed. Furthermore, the decomposition can be alternated along different dimensions, for instance to recover an alternating structure with gates along the $x$, $y$, etc dimensions at each length scale, as shown for 2D in \Fig{fig:QFT}~(b). 

So far, this gives an efficient prescription for performing the Fourier transform on fermionic systems in arbitrary dimensions using a log-depth circuit and two-body gates. This is potentially useful for state preparation or measurement schemes. More generally, a linear circuit with the same structure can be applied to bosonic systems. For bosons, $\hat{F}_2$ is the beam-splitter operation that maps two input annihilation operators $(\hat{a}_1,\hat{a}_2)$ to $(\hat{a}_1 + \hat{a}_2,\hat{a}_1 - \hat{a}_2)/\sqrt{2}$, while the phase gate $\hat{Z}^{a/b}$ transforms $\hat{a}_1$ to $(-1)^{a/b}\hat{a}_1$. Unlike fermions, the local Hilbert-space dimension is not fixed at two and in general can grow as large as the total number of particles in the system (and thus will not lead to a numerically efficient scheme as discussed below --- though this has already been suggested as a powerful tool in linear-optics quantum computation~\cite{Barak2007}). It would be interesting to determine if the structure is applicable to anyonic or spin models. 

\SectionCommand{The spectral tensor network}

\begin{figure}[t!]
\begin{centering}
  \includegraphics[width=1\columnwidth]{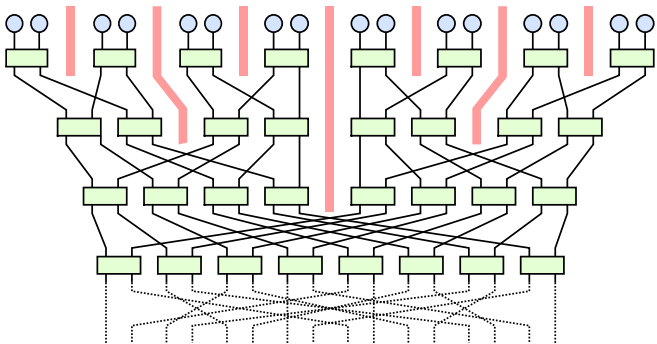}
  \caption{(Color online) The (inverse) Fourier transform on 16 sites applied to a product momentum-space state (blue circles) is redrawn as a spectral tensor network. At each layer, two-body unitaries couple two initially unentangled regions (emphasized by the wide red barriers).  At the bottom, the sites are reordered according to a bit-reversal permutation (dotted lines). \label{fig:STN}}
\end{centering}
\end{figure}

Consider a chain of $n = 2^k$ sites described by the second-quantization formalism, that begin in the simple Fock-state $|\Psi_0\rangle = |\alpha_1\rangle|\alpha_2\rangle\dots|\alpha_n\rangle = \hat{c}^{\dag \alpha_n}_n \dots \hat{c}^{\dag \alpha_1}_1 |0\rangle$. The circuit corresponding to the inverse Fourier transform can be applied to an (unentangled) product-state wavefunction in momentum space to yield a (highly-entangled) wavefunction in real space, as depicted in \Fig{fig:STN}. I call this tensor network structure a `spectral tensor network', and the wavefunction is a spectral tensor network state. In general, the connectivity of the diagram is that of a hypercube with side length 2 (even in higher spatial dimensions). Ignoring completely the permutation of sites at the bottom of \Fig{fig:STN}, the state resembles a branching MERA with half of the gates removed (making it significantly cheaper to contract), or similarly a generalized TTN. However, the site reordering turns the traditional concept of real-space renormalization on its head, as the tensor network couples sites with \emph{maximal} separation first. An intuitive explanation more likely comes from the decomposition of the cyclic group $C_n$ corresponding to translation invariance that motivates \Eqn{decompose} and the FFT (where ${C}_{pq} = {C}_p \times ({C}_{pq} / {C}_p)$, which is isomorphic to $C_p \times C_q$).

When performing calculations in the Fock basis, one must remember to respect the correct statistics of the degrees of freedom in the tensor network diagram by performing the appropriate permutation operation when two wires cross --- specifically, introducing a phase of $-1$ whenever an odd number of fermions cross~\cite{Barthel2009,Corboz2009,Corboz2009a,Kraus2009}. This phase can be accounted for by including an extra tensor at every crossing, fortunately without increasing the scaling of the computational cost. 

Together, this gives us a prescription for describing translationally-invariant free-particle systems in arbitrary dimensions, because these Hamiltonians are exactly diagonalized by the Fourier transform. In particular, we have a tensor-network wavefunction for free fermions in arbitrary dimensions. As is well known, such wavefunctions can exhibit quasi-long range correlations (polynomially decaying) and logarithmic violations to the area law. Nonetheless, we now show that extracting local expectation values from these wavefunctions can be achieved by an efficient tensor network contraction scheme, scaling (essentially) linearly with $n$.

\begin{figure}[t!]
\begin{centering}
  \includegraphics[width=1\columnwidth]{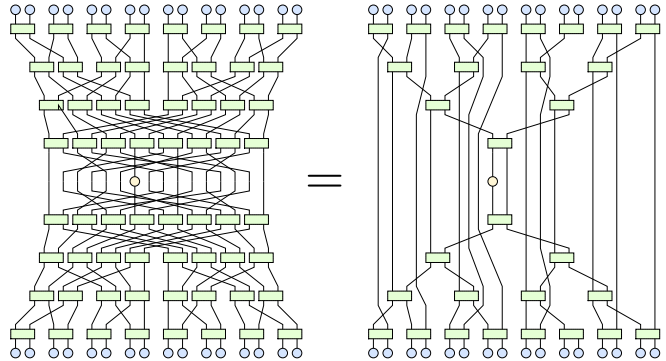}
  \caption{(Color online) The tensor network corresponding to the expectation value of a operator with support on a single site out of 16. Most of the tensors cancel due to unitarity, leaving a network that is contractible with cost $\mathcal{O}(\chi^5  n)$, where $\chi$ is the bond-dimension of the wires and $n$ is the number of sites.
  \label{fig:unitary_cancel}}
\end{centering}
\end{figure}

First, we take advantage of the unitary nature of the tensors to vastly simplify the calculation of expectation values of local operators. Figure \ref{fig:unitary_cancel} shows the tensor network corresponding to the expectational value of a single-site operator. The number of tensors reduces from $\mathcal{O}(n \log n)$ to $\mathcal{O}(n)$ and by contracting the remaining tree structure from the outside-in, the value can be calculated with computation cost $\mathcal{O}(\chi^5 n)$ (where $\chi$ is the dimension of the local Hilbert space). Full details of the contraction are found in the Appendix. Further, a set of one-site expectation values (or indeed, every one-site reduced density matrix) can be calculated with cost $\mathcal{O}(\chi^5 n \log n)$ by recycling previous calculations. Similarly, arbitrary two-site expectation values can be calculated with cost $\mathcal{O}(\chi^8 n)$ (see Appendix), and all two-body correlations in a translationally invariant system can be recovered with cost $\mathcal{O}(\chi^8 n \log n)$. For the systems we study here, both ground states and (grand-canonical) thermal states can be studied because the transformation exactly diagonalizes the Hamiltonian.

\begin{figure}[t!]
\begin{centering}
  \includegraphics[height=3.45cm]{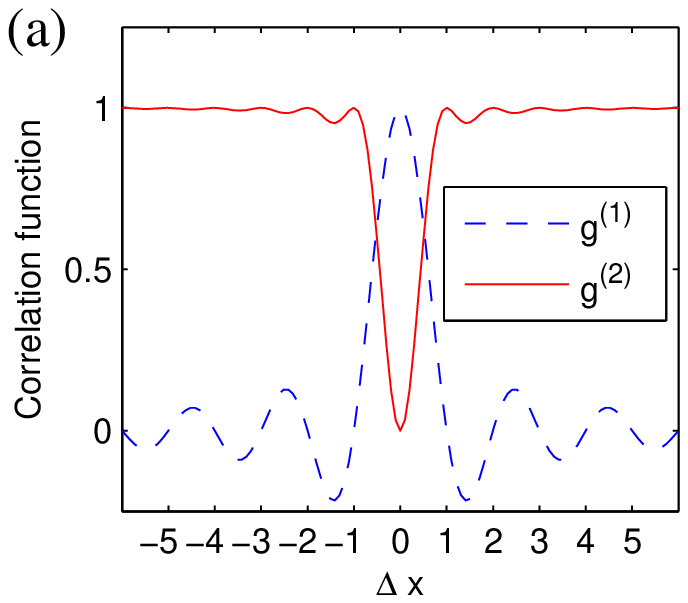}\includegraphics[height=3.45cm]{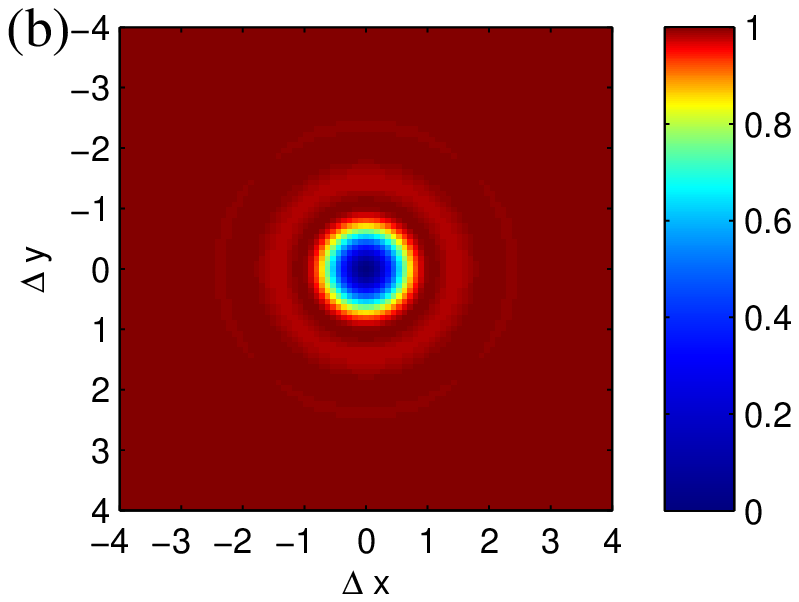}
  \caption{(Color online) (a)~The first-order and second-order normalized correlation functions $g^{(1)}$ and $g^{(2)}$ for the ground state of a 1D system of free fermions (103 fermions on 1024 sites). (b)~Second-order normalized correlation function $g^{(2)}$ for a 2D system of free fermions (2093 fermions on a $512 \times 512$ lattice). In both cases the distance ($\Delta x$ and $\Delta y$) is normalized to the mean particle spacing.
  \label{fig:correlations}}
\end{centering}
\end{figure}

As an simple example, we examine free-fermions in 1D and 2D with nearest-neighbour tunneling
\begin{equation}
    \hat{H} = -\sum_{<i,j>} \hat{c}^{\dag}_i \hat{c}_j + \mathrm{h.c.}
\end{equation}
where $<\!\!i,j\!\!>$ corresponds to neighboring sites on the lattice with $i < j$ (or similar in higher dimensions). We create the ground state wavefunction for a low filling factor and calculate the normalized first- and second-order correlation functions:
\begin{equation}
    g^{(1)} (\Delta x) = \frac{\langle \hat{c}^{\dag}_0 \hat{c}_{\Delta x} \rangle}{\langle \hat{n}_0 \rangle}, \; \; g^{(2)} (\Delta x) = \frac{\langle \hat{n}_0 \hat{n}_{\Delta x} \rangle}{\langle \hat{n}_0 \rangle^2}.
\end{equation}
The numerical results for 1D and 2D are shown in \Fig{fig:correlations}. These results should be considered exact, to numerical precision. The 2D simulation describes a highly-entangled system (in real-space) of more than quarter of a million sites, yet the calculation was easily completed on a laptop computer using MATLAB (of course, in this case the results could be obtained analytically or by alternative numerical techniques).

\SectionCommand{Generic free fermion systems}

The quantum circuit structure above can be used to construct translationally invariant states in a very wide range of phases by a straightforward extension to more spin or spatial degrees of freedom, multiple species, and/or including a Bogoliubov transformation to treat anomalous Hamiltonian terms (such as $\hat{c}_i \hat{c}_j + \mathrm{h.c.}$). 

In general, we can consider each site to allow $s$ different types of fermion, described by a Hilbert space of dimension $2^s$. Then the Fourier transform from a state initially unentangled in momentum space takes the same form as before, by treating each fermion type individually. The entire state can be encapsulated in a spectral tensor network with bond dimension $\chi = 2^s$. Conversely, the Fourier transform partially diagonalizes translationally invariant free-fermion models, resulting in a Hamiltonian that is local in momentum space (but couples the different species in a momentum-dependent fashion) --- meaning the spectral tensor network can describe all eigenstates of translationally invariant bilinear Hamiltonians. With $s>1$ we have a richer set of possibilities in a given momentum mode than simply `occupied' or `unoccupied', resulting in wider range of possible phases. These include the standard multiple-band possibilities including gapped insulators, standard gapless metals, higher-dimensional systems with interesting Fermi surfaces such as the Dirac point observed in graphene, and perhaps most interestingly, topological systems with non-trivial Chern numbers, as seen in the integer quantum Hall effect. These include as a subset the chiral PEPS states given in \Ref{Wahl2013}. 

Hamiltonians with anomalous terms are only partially diagonalized by the Fourier transform, leaving coupling between opposite momentum modes $\pm k$. The Supplemental Material describes how to include the Bogoliubov transformation into the spectral tensor network directly in momentum space while not changing the $n\log n$ computational scaling. This addition allows us to treat systems with superconductivity, Majorana fermion modes, or symmetry-protected topological order.

Some spectral tensor network ground states, such as insulators, chiral systems, and 2D systems with Dirac points have entanglement scaling with the area law. Meanwhile, other states such as those in the metallic phase contain significantly more entanglement with a logarithmic violation. It is even possible to achieve a volumetric scaling, for example, with a high-energy eigenstate or the ground state of a long-range Hamiltonian with random couplings. As far as the author is aware, the spectral tensor network is able to exactly describe a wider range of phases and dimensions than any other efficiently-scalable tensor network ansatz.

An additional result is a compact and exact tensor-network representation of strongly-correlated 1D systems that can be analytically solved using a Jordan-Wigner transformation. The Appendix provides details on the process and uses the transverse-field Ising model as an example. 

\SectionCommand{Simulation of arbitrary fermionic systems}


More generic systems are free to be studied by employing the variational method on the manifold of states defined by the tensor network diagram (in addition to the unitary constraints and any symmetry constraints of the model, e.g. the fermion parity). For instance, simple energy minimization may yield states very close to highly-entangled ground states, particularly in the Fermi-liquid phase where the entanglement structure should be related to that of free fermions. The bond-dimension of the tensor network can be increased by combining neighboring lattice sites, which increases the fraction of Hilbert space covered by the ansatz and thus potentially the accuracy of the minimum energy state. This is an area of intense interest to the author and is the subject of further work. Furthermore, it would be interesting to see if the spectral tensor network can accurately represent systems with different boundary conditions or impurities, or be applied effectively to spin systems.

\SectionCommand{Conclusion}

The spectral tensor network has been introduced here as a way of efficiently representing the quantum fast Fourier transform and certain classes of quantum states. The Fourier transform can be used to construct free-fermion states with a logarithmic violation of the area law, and as far as the author is aware is the first efficiently contractible tensor network that has been shown to \emph{exactly} capture this phase in two- and higher-dimensions.

The tensor network is shown to be a convenient form for capturing analytic solutions to 1D spin problems that are solved by the Jordan-Wigner transformation. Further, we suggest that this tensor network may be used as a variational ansatz for low-energy states of many-body systems. This may be especially effective in the case of Fermi-liquid phases, where from a \emph{local} perspective the system is highly entangled and is somewhat difficult to treat with existing tensor network techniques, while the Fourier transform is able to convert the system into a relatively disentangled momentum-space basis. Work in this direction is ongoing.

The spectral tensor network is somewhat related to the branching MERA construction of Evenbly and Vidal, with two major differences. The spectral tensor network contains only half the tensors of the branching MERA and is more efficient numerically. A more striking difference is the non-local reordering of lattice sites placed at the bottom of the unitary circuit. In complete opposition to the spirit of real-space renormalization, the unitaries at the bottom of the circuit deal with the \emph{longest} length-scales of the system, with the length scales \emph{decreasing} as we ascend the diagram. 

\subsection{Acknowledgements}

I would like to thank Guifr\'e Vidal and Luca Tagliacozzo for stimulating discussions. This work was supported by TOQATA (Spanish grant PHY008-00784), the EU IP SIQS, the MPQ-ICFO collaboration, and NSERC and FQRNT through the network INTRIQ.

\bibliography{../bib/andy}

\section{Appendix}

Here we give additional technical details to complement the main text. We begin by giving a fuller derivation of the quantum fast Fourier transform for fermion/bosonic systems, and discuss the inherent freedom allowed by the decomposition. We then detail the tensor contractions necessary for performing calculations with the spectral tensor network, including derivations of the computational cost of measuring one- and two-site local variables and performing ground state optimizations in arbitrary dimensions. We also consider the extended case corresponding to a Fourier transform combined with a Bogoliubov transform that couples modes with opposite momentum, which should be contracted in a slightly different order to remain efficient. The 1D transverse-field Ising model is given as an example application of how both the Jordan-Wigner transformation and Bogliubov transform can be used in conjunction with the spectral tensor network to efficiently extract properties of strongly-correlated systems.

\subsection{Derivation of the quantum fast Fourier transform}

The Fourier transform for fermions (and bosons) over $n$ (assumed even) sites is given by the following linear transformation:
\begin{equation}
  \tilde{c}^{\dag}_k = \frac{1}{\sqrt{n}} \sum_{x=0}^{n-1} e^{\frac{2 \pi i k x}{n}} \hat{c}^{\dag}_x .
\end{equation}
We now define two subsystems containing the even sites:
\begin{equation}
   \hat{a}^{\dag}_x = \hat{c}^{\dag}_{2x},
\end{equation}
and the odd sites:
\begin{equation}
   \hat{b}^{\dag}_x = \hat{c}^{\dag}_{2x+1}.
\end{equation}
Each subsystem contains $n/2$ sites and have associated Fourier transforms:
\begin{equation}
  \tilde{a}^{\dag}_k = \frac{1}{\sqrt{n/2}} \sum_{x=0}^{n/2-1} e^{\frac{2 \pi i k x}{n/2}} \hat{a}^{\dag}_x ,
\end{equation}
\begin{equation}
  \tilde{b}^{\dag}_k = \frac{1}{\sqrt{n/2}} \sum_{x=0}^{n/2-1} e^{\frac{2 \pi i k x}{n/2}} \hat{b}^{\dag}_x .
\end{equation}

It is straightforward to verify the following relation (Eq. (1) of the main text), which decouples even and odd sites in the full $n$-site Fourier transform:
\begin{equation}
  \sum_{x=0}^{n-1} e^{\frac{2 \pi i k x}{n}} \hat{c}^{\dag}_x = \!\sum_{x^{\prime} = 0}^{n/2-1} \!e^{\frac{2 \pi i k x^{\prime}}{n/2}} \hat{c}^{\dag}_{2x^{\prime}} + e^{\frac{2 \pi i k}{n}} \!\sum_{x^{\prime} = 0}^{n/2-1} \!e^{\frac{2 \pi i k x^{\prime}}{n/2}} \hat{c}^{\dag}_{2x^{\prime}+1} \label{decompose_repeat} 
\end{equation}
This can be written in terms of the above variables as
\begin{equation}
\tilde{c}^{\dag}_k = \frac{\tilde{a}^{\dag}_k + e^{\frac{2 \pi i k}{n}}  \tilde{b}^{\dag}_k}{\sqrt{2}}, \label{twobodytransform}
\end{equation}
where we note that $\tilde{a}^{\dag}_{k+n/2} = \tilde{a}^{\dag}_k$ and similarly $\tilde{b}^{\dag}_{k+n/2} = \tilde{b}^{\dag}_k$. Equation \ref{twobodytransform} is a simple two-body, linear transformation between modes $\{\tilde{a}^{\dag}_k,\tilde{b}^{\dag}_k\}$ and $\{\tilde{c}^{\dag}_k,\tilde{c}^{\dag}_{k+n/2}\}$. Generically this transformation can be represented with the unitary $(\hat{\mathbb{1}} \otimes\hat{\omega}_{n}^{k}) . \hat{F}_2$, where
\begin{equation}
  \hat{F}_2^{\dag} \, \tilde{c}^{\dag}_k \, \hat{F}_2 = \frac{\hat{c}^{\dag}_k  + \hat{c}^{\dag}_{k+n/2}}{\sqrt{2}}
\end{equation}
\begin{equation}
  \hat{F}_2^{\dag} \, \tilde{c}^{\dag}_{k+N/2} \, \hat{F}_2 = \frac{\hat{c}^{\dag}_k  - \hat{c}^{\dag}_{k+n/2}}{\sqrt{2}}
\end{equation}
\begin{equation}
  \hat{\omega}_n^{k\dag} \, \tilde{c}^{\dag}_{k+n/2} \, \hat{\omega}_n^k = \exp\left(\frac{2\pi i k}{n}\right) \hat{c}^{\dag}_{k+n/2}
\end{equation}

This allows us to decompose the Fourier transform over $n$ sites into smaller gates, which include the two smaller Fourier transforms for $n/2$ the $\hat{a}^{\dag}_x$ and $\hat{b}^{\dag}_x$ modes (i.e. the even and odd sites), as well as $n/2$ one- and two-body linear gates. By iterating the decomposition recursively, the remaining Fourier transforms can decomposed into one- or two-body gates. The complete decomposition has depth $\sim \log_2(n)$ with a total of $\sim n\log_2 n$ gates.

\begin{figure}[t!]
\begin{centering}
  \includegraphics[width=1\columnwidth]{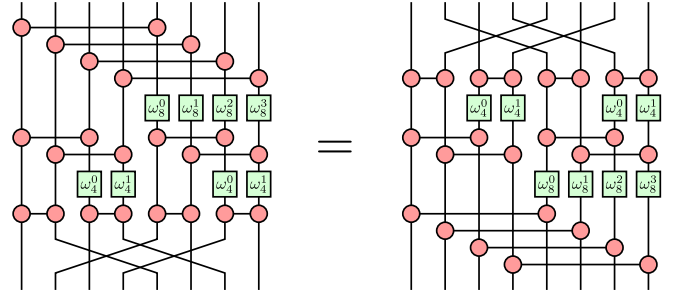}
  \caption{(Color online) Unitary circuit for performing the quantum Fourier transform on 8 fermionic sites, $\hat{F}_8$. The red gates are the two-site Fourier transform, $\hat{F}_2$. The layers can be permuted, resulting in different `twiddle factors' $\hat{\omega}^a_b$ and reorderings. The reordering of the sites at the top or bottom of the circuit is given by the bit-reversal operation. \label{fig:QFT2}}
\end{centering}
\end{figure}

\begin{figure}[t!]
\begin{centering}
  \includegraphics[width=1\columnwidth]{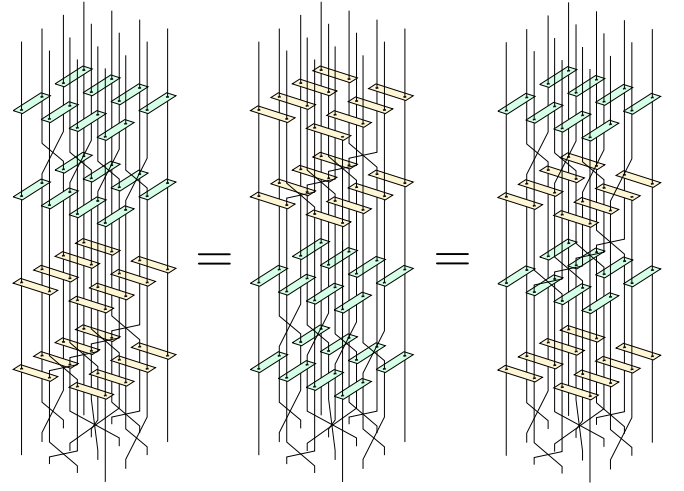}
  \caption{(Color online) Unitary circuit for performing the Fourier transform on a 2D square lattice system of $4 \times 4$ sites. Here we have absorbed the twiddle factors into the two-body gates, thus each gate depicted is unique. There is significant freedom in choosing the order of the transformation layers.
  \label{fig:QFT_2D}}
\end{centering}
\end{figure}

One interesting feature of the fast Fourier transform is the flexibility of in the decomposition, which comes from three sources: (a) graphical manipulations of the unitary circuit/tensor network diagram, (b) the fact that the Fourier transform is self-transpose, $\hat{F}_n = \hat{F}_n^{T}$, and (c) commutability between different dimensions. Graphical manipulations include `pushing' the permutation through to the top of the diagram, which simply rearranges the gates and means the permutation performed at the momentum space side of the diagram. The self-transpose property can be applied at different stages of the decomposition. An example using 8 sites is provided in \Fig{fig:QFT2}. Furthermore, in two or higher dimensions the transformation layers along \emph{different} directions commute --- see for example \Fig{fig:QFT_2D}. This flexibility is analogous to that available to numerical algorithms for the classical FFT.

As a final note, it should be possible to extend this decomposition to system sizes that are not a power of two while retaining the efficiency. For instance, the decomposition works for factoring out of $n$ any prime factor $p$, but would necessitate $p$-body gates and therefore increase the computation scaling as a power of $\chi$. The classical Fourier transform can calculated efficiently (i.e. with $n \log n$ scaling) even for large primes $n$~\cite{Bluestein1968}, and it would be interesting to determine if a simple quantum circuit can be deduced for these situations, and furthermore whether it would be associated with an efficiently contractible tensor network. However this is beyond the scope of the present work.

\subsection{Tensor contractions for one- and two-site operators, and optimization}

\begin{figure}[t]
\begin{centering}
  \includegraphics[width=1\columnwidth]{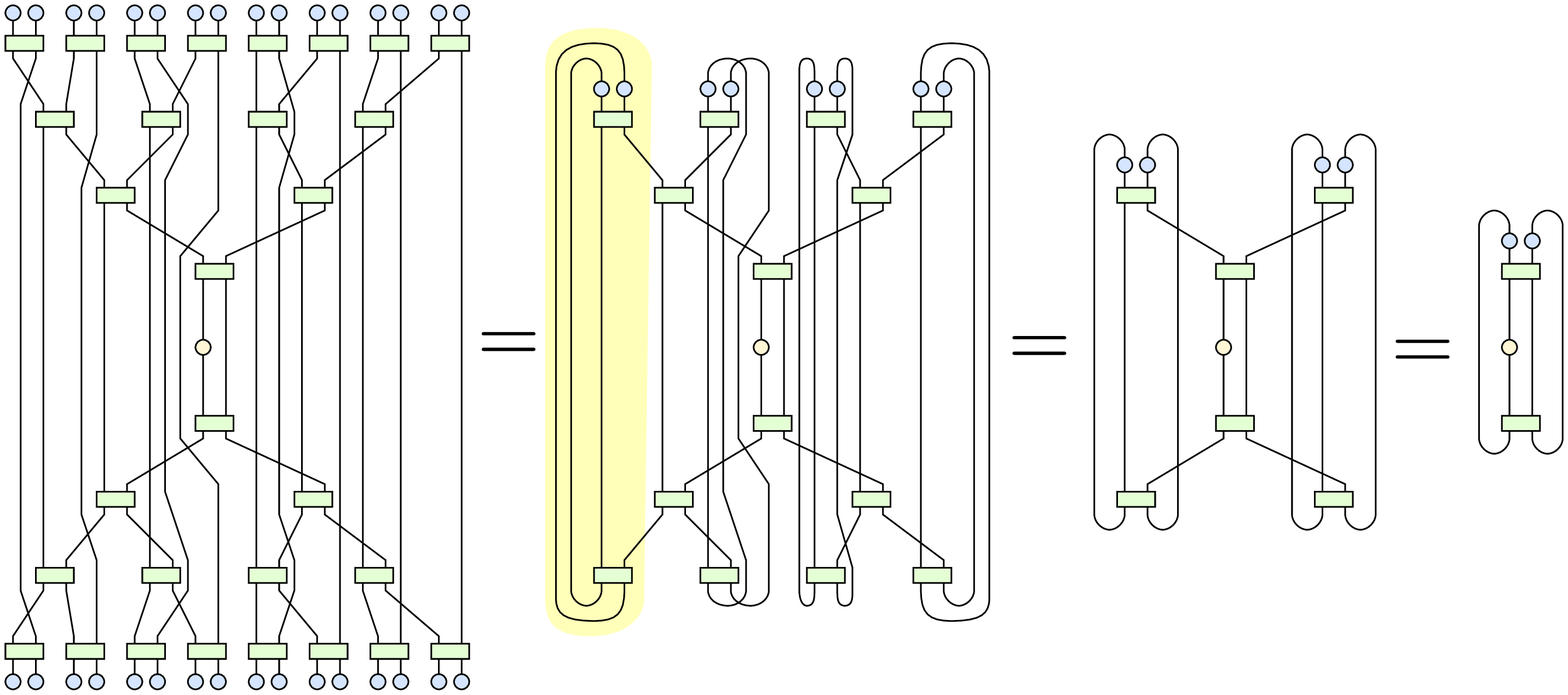}
  \caption{(Color online) The expectation value of a single-body operator can be contracted efficiently by contracting indices on the outside of the diagram and working inwards. Each primitive step consists of contracting a section of the diagram like that highlighted in yellow, corresponding to the operation in \Fig{fig:steps}~(a). This has to occur $n$ times, leading to a total cost scaling as $\mathcal{O}(n\chi^5)$. 
  \label{fig:onesitecontraction}}
\end{centering}
\end{figure}

In the main text, we briefly showed that the expectation value of a single-site observable can be calculated with cost $\mathcal{O}(n)$. To reiterate this more precisely, we describe further both the process of removing unitary gates from the expectation value and the contraction scheme required to efficiently compute the final value.

First, we must recall that the tensor network diagrams here are for fermionic Hilbert spaces. Each bond (or wire) in diagrams such as Fig.~3 of the main text represent one or more fermionic modes. The tensor network language lets us graphically express the total Hilbert space \emph{as if it were a product of independent Hilbert spaces} for each mode (or wire). However, we know such a decomposition is \emph{not} possible for fermionic Hilbert spaces, at least in the usual sense, and thus additional rules are required to remain consistent with fermionic statistics. Specifically, when two wires cross, this is equivalent to algebraically re-ordering our fermionic modes and thus should introduce a factor of $-1$ when an odd number of fermions pass through each wire. One way of encoding this is to place a diagonal operator ("mini-tensor") at each of the wire crossings, and then the resulting diagram can be interpreted and computed in the usual (bosonic) sense.

Furthermore, it is always possible to \emph{choose} a new ordering of the fermionic modes that we use to define our Fock-space basis, in either the traditional algebraic or graphical tensor network language. In the diagrams, this is achieved by dragging the wires over each other and pulling them through tensors, each of which is allowed without changing the result (note that mini-tensors at wire crossings may be added or removed because of these manipulations). These graphical rules are a powerful tool for simplifying tensor network diagrams. As an example, consider Fig.~3 of the main text. In this equation, all of the unitaries that are not causally connected to the operator at the 7th site cancel. To see this cancellation clearly, we can drag around the wires in the centre (those that cross the "physical" plane) so that the cancelling unitaries multiply their conjugates without wire crossings in-between (at which point they can be removed from the diagram). In some cases (e.g. tensors further from the centre) it may even be necessary to move the unitary tensors up or down to achieve this effect. The freedom given by the allowed manipulations leads to the conclusion that wire crossings do not impair the application of the unitarity identity $\hat{U} \hat{U}^{\dag} = \hat{I}$.

\begin{figure}[t]
\begin{centering}
  \includegraphics[width=0.9\columnwidth]{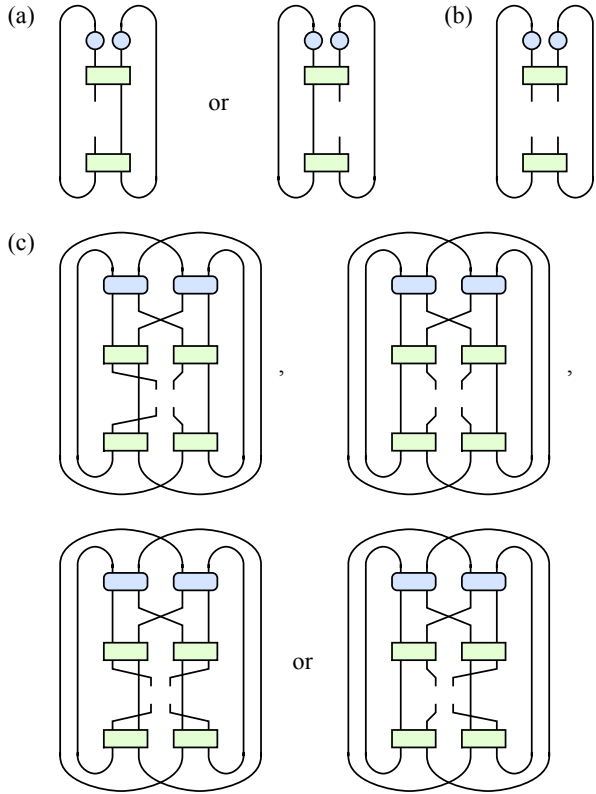}
  \caption{(Color online) The scaling operators necessary to contract expectation values of one- and two-body operators. The open wires are arranged to make the necessary fermionic wire-crossing operations more obvious. (a) Two single-site density operators (blue) are transformed into a new single-site operator, with open indices either in the left or right `direction' (cost $\mathcal{O}(\chi^5)$. (b) Two single-site density operators are transformed (unitarily) into a new two-site operator. (c) Two two-site density operators are transformed into a single two-site operator, with four possible descending directions (cost $\mathcal{O}(\chi^5)$. Where the wires cross, a phase of $-1$ is introduced when an odd number of fermions are contained in each wire (this could be represented with the addition of diagonal operators in the diagram, having no impact on the leading-order computation cost). Note how in each case the four open indices are brought to the same geometric region, so that a well-defined tensor for a fermionic operator results (otherwise it is easy to loose track of the correct mode-ordering when combining tensors in later iterations).
  \label{fig:steps}}
\end{centering}
\end{figure}

Next we consider the contraction scheme for efficiently computing this expectation value, as depicted in \Fig{fig:onesitecontraction}. The contraction proceeds by working from the outside inwards, taking advantage of the very small amount of connectivity in the diagram (consider that the diagram folded around a horizontal line through the center is a simple tree). Each of these basic steps follows the diagrams in \Fig{fig:steps}~(a) and have cost $\mathcal{O}(\chi^5)$, and occur $n$ times, leading to a total cost of $\mathcal{O}(\chi^5 n)$.

\begin{figure}[t]
\begin{centering}
  \includegraphics[width=1\columnwidth]{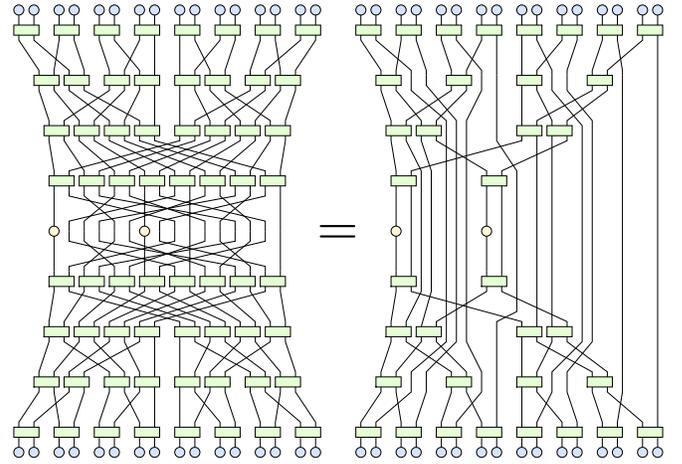}
  \caption{(Color online) The tensor network contraction for computing the expectation of a two-body operator. Many of the tensors have cancelled due to unitarity, and the remaining can be contracted with cost $\mathcal{O}(n\chi^8)$.
  \label{fig:twosite}}
\end{centering}
\end{figure}

The expectation value of few-site observables can also be calculated efficiently with cost $\mathcal{O}(n)$, although with a cost which grows exponentially with the size of the operator in question. In \Fig{fig:twosite}, the necessary contraction for a widely-separated, two-site operator is shown. Note here that we are really working in the fermionic Hilbert space and it is unnecessary to include a string of operators between the two observables, as would be typically seen in the Jordan-Wigner transformation (this is a demonstration of the power and simplicity of the graphical, fermionic tensor network language). Like we saw earlier, many of the unitaries cancel each other resulting in a sparser tensor network diagram.

As before, this diagram can be contracted using simple steps starting from the outside of the diagram and working towards the physical indices, layer-by-layer. These steps each cost up to $\mathcal{O}(\chi^8)$ for two-site operators, and are shown in \Fig{fig:steps}. At the outer-layers, the single-site contractions (\Fig{fig:steps}~(a)) suffice until the two sites `fuse' in the tree, described by \Fig{fig:steps}~(b). After this point, a effective two-site `density operators' must be carried through the diagram as in \Fig{fig:steps}~(c). For a given observable, $n$ such steps are necessary, making the total cost $\mathcal{O}(\chi^8 n)$ for a two-site observable. The exponent of $\chi$ would grow for observables over more sites.

Fortunately, when computing a range of observables many computations can be reused to reduce the total cost. For instance, in a non-translationally invariant model, the entire set of one-body observables $\langle \hat{A}_i \rangle$ can be calculated with cost $\mathcal{O}(\chi^5 n \log n )$. This is achieved by noting that, over all values of $i$, only two versions of the outside contraction is needed and can be reused (corresponding to the left and right versions of the diagrams in \Fig{fig:steps}~(a)). However, there are $n/2$ two-body unitaries in that layer of the circuit, leading to $n$ unique calculations. At the next layer, there are now four possible `directions' to account for but only $n/4$ unitaries, so there are again $n$ unique calculations to perform --- and so on for every layer. Because there are $\log_2 n$ layers, the total number of unique contraction steps required is reduced to just $n \log_2 n$ , while a na\"ive scheme would perform $n^2$ contraction steps.

As another example of this speed-up, in a translationally-invariant system, the entire set of two-body observables $\langle \hat{A}_0 \hat{B}_i \rangle$ can be calculated with cost $\mathcal{O}(\chi^8 n \log n )$ (as was done to generate Fig. 5 in the main paper). The same scaling applies to the expectation value of local, two-body Hamiltonians (with no assumption on translational invariance).

With a simple modification to the above procedure, one can additionally recover the derivative of the expectation value with respect to the tensors themselves (the so-called `environment' of each tensor) with the same $\mathcal{O}(\chi^8 n \log n)$ scaling. This allows us to follow a derivative-based optimization procedure for determining the state with minimum energy, such as steepest-descent within the unitary subspace~\cite{Ferris2011b}, singular-value decomposition updates~\cite{Evenbly2007} or more complicated approaches~\cite{Haegeman2011}.

\subsection{Jordan-Wigner transformation for spin chains and Bogoliubov transformations}

\begin{figure}[t]
\begin{centering}
  \includegraphics[width=0.5\columnwidth]{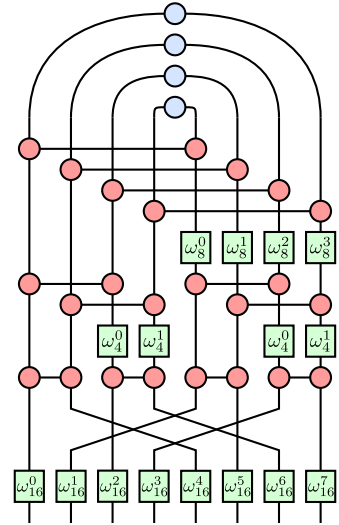}
  \caption{(Color online) A spectral tensor network state with additional links between $\pm k$ momentum modes and `half-integer' momenta. The eigenstates of translationally-invariant, quadratic Hamiltonians with anomalous terms can be represented with this tensor network.
  \label{fig:bog}}
\end{centering}
\end{figure}

\begin{figure}[t]
\begin{centering}
  \includegraphics[width=1\columnwidth]{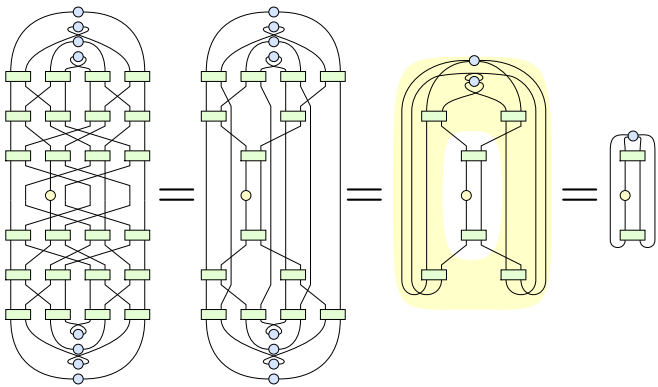}
  \caption{(Color online) Example calculation of a single-site expectation value with a Bogoliubov state. In this diagram, some of the Bogliubov tensors at the top have been flipped horizontally compared to \Fig{fig:bog}. Efficient contraction relies on unitarity (first equality) followed by contracting the diagram from the outside in. The typical step for large systems is highlighted in yellow, corresponding to first diagram in \Fig{fig:bogsteps}.
  \label{fig:bogcontraction}}
\end{centering}
\end{figure}

The spectral tensor network is also useful in treating non-fermionic systems. For example, the Jordan-Wigner transformation can be used to transform some 1D spin systems into non-interacting fermionic models. A straight-forward example is the fully-anisotropic Heisenberg spin-1/2 chain (XX model), which transforms into the free-fermion system studied in the main paper. Other models, such as the quantum transverse-field Ising chain on $n$ sites with Hamiltonian
\begin{equation}
    \hat{H} = \sum_i \hat{X}_i \hat{X}_{i+1} + h \hat{Z}_i .
\end{equation}
are slightly more complicated because the require a Bogoliubov transformation between opposite momentum modes. This system corresponds to a free fermion system by the transformation $\hat{c}_i = \frac{1}{2}(\hat{X}_i + i\hat{Y}_i) \prod_{j<i} \hat{Z}_j$ where $\hat{c}_i$ are fermion annihilation operators, resulting in
\begin{equation}
    \hat{H} = 2\sum_i (\hat{c}^{\dag}_i - \hat{c}_i) (\hat{c}^{\dag}_{i+1} + \hat{c}_{i+1}) + h \hat{c}_i^{\dag} \hat{c}_i - 2n.
\end{equation}
This Hamiltonian is bilinear in the fermionic operators and translationally invariant, and so can be solved using a Fourier transform. In the case of a quadratic Hamiltonian with anomalous pair terms, like above, a Bogoliubov transformation coupling modes with opposite momentum will diagonalize the Hamiltonian. That is, we couple $\hat{\tilde{c}}_k$ with $\hat{\tilde{c}}^{\dag}_{-k}$, where $\hat{\tilde{c}}_k = \sum_j \exp(2\pi i j k) \hat{c}_j/\sqrt{n}$, by applying an additional transformation at the `top' of the tensor network diagram, as shown in \Fig{fig:bog}.

\begin{figure}[t]
\begin{centering}
  \includegraphics[width=0.78\columnwidth]{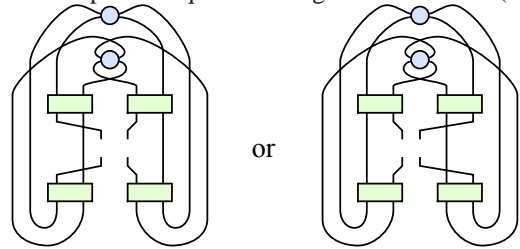}
  \caption{(Color online) The scaling operators necessary to contract expectation values of one-site operators with the Bogoliubov transformation, each costing $\mathcal{O}(\chi^8)$.
  \label{fig:bogsteps}}
\end{centering}
\end{figure}

However, contracting the tensor network na\"ively in the previous order will result in growing bond-dimensions in the intermediates steps. To perform the contraction efficiently, we simply perform the layer-by-layer `renormalization' on the coupled pairs instead of the individual sites, as shown for a single-site observable in \Fig{fig:bogcontraction}. At each step, we combine two pairs of sites with a unitary transform and partial trace, resulting in a density matrix for a single pair of sites. We can see the steps required for calculating single-site observables in \Fig{fig:bogsteps}, costing $\mathcal{O}(\chi^8)$ each. Like we saw previously, $\mathcal{O}(n)$ such steps are required for a given observable (and just $\mathcal{O}(n \log n)$ steps for a full set on every site by reusing calculations). A similar procedure applies to the case of observables over two or more sites.

\begin{figure}[t]
\begin{centering}
  \includegraphics[width=0.8\columnwidth]{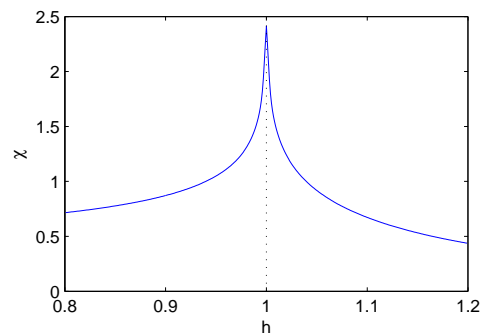}
  \caption{(Color online) Numerical results for magnetic susceptibility the transverse-field Ising model on a lattice of 1024 sites around the critical point at $h = 1$. 
  \label{fig:TFI}}
\end{centering}
\end{figure}

We have implemented and tested this procedure for the transverse-field Ising model. In \Fig{fig:TFI} we present results for the magnetic susceptibility in the field direction, based on a simple numerical derivative evaluating
\begin{equation}
   \chi = -\frac{d \langle \hat{Z} \rangle}{dh}.
\end{equation}
We observe a sharp peak at the critical point at $h=1$, as expected.

The scheme presented here might be improved on in several ways. For instance, it has not yet been investigated if the spectral tensor network can represent the Bogoliubov transformation natively, that is, without the extra tensors at the top linking $\pm k$ modes, which is the subject of future work.

\end{document}